\documentclass[11pt]{article}

\usepackage[a4paper,margin=1in]{geometry}
\usepackage[T1]{fontenc}
\usepackage[utf8]{inputenc}
\usepackage{lmodern}
\usepackage{amsmath,amssymb,amsfonts}
\usepackage{booktabs}
\usepackage{array}
\usepackage{multirow}
\usepackage{graphicx}
\usepackage{xcolor}
\usepackage{hyperref}
\usepackage[numbers]{natbib}
\usepackage{authblk}
\usepackage{enumitem}
\usepackage{caption}
\usepackage{longtable}
\usepackage{pdflscape}
\usepackage{algorithm}
\usepackage{float}
\usepackage{xspace}
\usepackage{algpseudocode}

\hypersetup{colorlinks=true,linkcolor=blue,citecolor=blue,urlcolor=blue}

\newcommand{\asymbol}{\ensuremath{\alpha}\xspace}

\title{\texorpdfstring{The $\alpha$-Index}{The alpha-Index}: A Penalized Authorship-Integrity Framework for Position-Weighted Scientific Contribution}

\author[1,2]{Athanasios Angelakis}
\affil[1]{BioML Lab, RI CODE, UniBw, Munich, Germany}
\affil[2]{Epidemiology and Data Science, Amsterdam UMC, Amsterdam, Netherlands}
\date{}

\begin{document}
\maketitle
  
\begin{abstract}
Publication and citation indicators commonly assign full credit to every coauthor, obscuring differences in authorship role and potentially rewarding accumulated authorship rather than identifiable intellectual contribution. We propose the $\alpha$-index, denoted by the Greek lowercase letter \asymbol, as a conserved, position-weighted, and penalized authorship-integrity framework. Each publication contributes one unit of credit, allocated across first-author execution, senior-author leadership, and residual middle authorship. Its defining feature is a senior-author responsibility penalty: senior credit decreases as the residual middle-author list expands, expressing the normative principle that leadership credit should be accompanied by responsibility for authorship discipline. The paper formalizes local allocation and a cumulative penalized global index; presents a parameterized family of weight blocks and penalty functions; and compares the framework with fractional, harmonic, and $h_{\alpha}$-type approaches. Synthetic examples and selected public byline illustrations demonstrate mathematical behavior, including large-team variants. The default values are not empirical constants but transparent, testable hypotheses within a calibratable family. The framework is presented as a methodological and ethical proposal requiring field-specific validation against contribution statements, expert assessments, author surveys, and bibliographic data. It is intended to complement, not replace, peer review, contributor statements, acknowledgements, and citation-based metrics.
\end{abstract}


\section{Introduction}

The h-index combines publication volume and citation impact into a single number and is now deeply embedded in academic evaluation \citep{Hirsch2005}. However, the h-index does not distinguish a sole-author paper from a first-author paper, a last-author paper, or a middle-author paper in a large author list. A researcher can increase conventional publication and citation-based metrics through repeated coauthorship even when their role is peripheral, routine, administrative, network-based, or weakly connected to the intellectual core of the work. This limitation is not only mathematical. It affects hiring, promotion, funding, reputation, and the perceived identification of scientific leadership.

The present paper is motivated by a simple concern: modern evaluation systems often reward the quantity of indexed authorships more than the quality and necessity of scientific contribution. In many biomedical and computational areas, the people most visibly responsible for the scientific work are usually the first or equal-first authors, who carry the main execution, and the last or equal-last authors, who carry senior intellectual leadership and accountability. Middle authors may certainly make important contributions, but the middle-author region is also where routine technical support, access provision, reciprocal authorship, administrative involvement, or network-based inclusion can be difficult to distinguish from genuine intellectual necessity. Historically, acknowledgements were used to recognize important but non-authorial support. In many contemporary publications, acknowledgement-like contributions appear increasingly to be absorbed into expanding middle-author lists.

International authorship recommendations, including the International Committee of Medical Journal Editors criteria, emphasize substantial contribution, accountability, drafting or critical revision, and final approval \citep{ICMJE2024}. Contributor taxonomies such as CRediT provide valuable qualitative role labels \citep{Brand2015}. Yet these frameworks do not provide a conserved numerical authorship-credit coefficient that can be compared across publication records. Existing bibliometric alternatives, including equal fractional counting, harmonic counting, harmonic publication and citation counting, and coauthorship-corrected h-type indices, address parts of the problem \citep{Batista2006,Wan2007,Schreiber2008,Hagen2008,Hagen2010,Vavrycuk2018,Sundling2023}. However, they do not directly encode the combined first-author execution principle, last-author leadership principle, and senior-author responsibility for authorship discipline.

We therefore propose the $\alpha$-index. The official $\alpha$-index is the penalized form of the framework. It is developed through three layers. First, the \emph{local penalized $\alpha$-credit} assigns a conserved paper-level credit value to each author. Second, the \emph{global $\alpha$-index} is defined as the cumulative sum of these penalized local $\alpha$-credits across an author's publication record. Third, the final $\alpha$-allocation applies a senior-author penalty that reduces senior-author credit as the number of residual middle authors increases. The purpose is not to deny legitimate team science. Rather, it is to make explicit a principle that current metrics largely ignore: if senior authors receive credit for leading a project, they should also bear bibliometric responsibility for controlling unnecessary authorship expansion. In the remainder of the manuscript, the term \emph{$\alpha$-index} refers to the default penalized authorship-integrity allocation. The unpenalized allocation is retained only as a transparent baseline. The paper should be read as a formal proposal and calibration framework, not as an empirically validated universal metric.

\section{Design principles}

A useful authorship-integrity metric should satisfy the following requirements.

\begin{enumerate}[label=R\arabic*.]
  \item \textbf{Credit conservation.} Each paper should contribute exactly one unit of authorship credit. One paper should not generate $n$ full credits simply because it has $n$ authors.
  \item \textbf{Execution recognition.} First and equal-first authors should receive the dominant share of credit in fields where first authorship reflects primary intellectual and technical execution.
  \item \textbf{Leadership recognition.} Last and equal-last authors should receive credit for senior intellectual leadership, supervision, accountability, and project direction.
  \item \textbf{Authorship-discipline penalty.} Senior-author credit should decrease when the middle-author list expands, because project leaders are responsible for distinguishing necessary intellectual contributors from acknowledgement-level support.
  \item \textbf{Local and global interpretability.} The metric should be meaningful at the paper level and aggregatable at the researcher level.
  \item \textbf{Transparency.} The rule should be simple enough to audit, reproduce, criticize, and recalibrate.
\end{enumerate}

The $\alpha$-index framework is normative. It does not claim that author order perfectly represents contribution in every field, nor that the default constants below are natural constants. Instead, it proposes a transparent and auditable default within a family of position-weighted authorship-integrity indices. The default is intended for fields where first and last authorship are meaningful and where authorship inflation is a recurring concern; other fields may require empirical calibration or a different member of the same family.

\section{A parameterized family rather than universal constants}
\label{sec:family}

A central revision motivated by methodological criticism is that the $\alpha$-index should be presented as a \emph{family} of conserved authorship-integrity allocations, with a clearly declared default. Let
\[
F+M+L=1,\qquad F,M,L\geq 0,
\]
where $F$ denotes the first-author execution block for papers with at least three authors, $M$ the residual middle-author block, and $L$ the senior-author leadership block. For two-author papers, let $F_2$ denote the first-author block and $1-F_2$ the second-author block. The default used in this paper is
\begin{equation}
F_2=0.67,\qquad F=0.67,\qquad M=0.10,\qquad L=0.23.
\label{eq:default_blocks_family}
\end{equation}
These values encode a normative stance: approximately two thirds of the local credit is assigned to the first-author execution role; a small residual block is reserved for middle authorship; and the remaining credit is assigned to senior leadership, subject to the penalty introduced below. The separate two-author parameter $F_2$ is included because the two-author case has no residual middle-author region and therefore does not fit the three-block structure exactly. The default sets $F_2=F$ for simplicity, but empirical calibration could choose $F_2\neq F$ if two-author authorship practices differ from multi-author practices.

The default values are not claimed to have been empirically derived. They should be read as testable hypotheses and a transparent starting point because they are simple, interpretable, and deliberately asymmetric in favor of identifiable execution. A fully validated index would require field-specific calibration using contribution statements, expert panels, author surveys, or observed research-evaluation decisions. The purpose of this manuscript is therefore to formalize the framework and examine its mathematical behavior, not to assert that one fixed triplet $(F,M,L)$ or one fixed value of $F_2$ is universally correct.

The two-author default $(0.67,0.33)$ is therefore not a claim that the second author in a two-author paper is equivalent to a last author in a larger biomedical or computational byline. It is a pragmatic default for the special case in which there is no middle-author block to allocate and no senior-author penalty to apply. In fields where two-author papers usually reflect equal intellectual partnership, $F_2$ may be set closer to $0.50$; in fields where the first-author convention is very strong, $F_2$ may be set closer to the default.

For any penalty function $g(m)$ satisfying $0<g(m)\leq 1$ and $g(1)=1$, the general penalized family can be written as
\begin{equation}
L^{P}_{g}(m)=L\,g(m),
\end{equation}
\begin{equation}
M^{P}_{g}(m)=1-F-L\,g(m),
\end{equation}
for $m>0$. A useful penalty function should satisfy three minimal properties: $g(1)=1$ so that a paper with one residual middle author reproduces the baseline senior block; $g(m)$ should be non-increasing so that senior credit does not increase as the residual middle-author list expands; and $g(m)$ should remain positive so that senior authorship is not reduced to zero for finite $m$. The choice among hyperbolic, square-root, logarithmic, or capped functions expresses how strongly a field wants to encode senior-author responsibility for byline expansion. The default $\alpha$-index uses the strict hyperbolic penalty $g(m)=1/m$. Softer functions, including $1/\sqrt{m}$ and $1/\log_2(m+1)$, are considered later as large-team variants. This parameterized presentation directly separates the conserved framework from the empirical question of which parameters are most appropriate for a given discipline.

\section{Authorship ethics and research-integrity motivation}

Authorship is not merely a mechanism for distributing academic reward. It is also a mechanism of accountability. Medical-journal guidance explicitly links authorship to substantial contribution, drafting or critical revision, approval of the final work, and responsibility for the integrity of the work \citep{ICMJE2024}. Publication-ethics guidance similarly treats gift authorship, guest authorship, ghost authorship, and unclear contribution standards as research-integrity problems rather than only technical bibliometric problems \citep{COPE2019,COPEGift}. Empirical studies have documented honorary and ghost authorship in biomedical journals, showing that questionable attribution is not only hypothetical \citep{Flanagin1998,Wislar2011}.

The ethical concern addressed here is not that every middle author is illegitimate. Many middle authors make essential conceptual, experimental, clinical, statistical, computational, or interpretive contributions. The concern is that the middle-author region can mix authorship-level contribution with acknowledgement-level support, such as routine technical processing, infrastructure access, recruitment, administrative support, network membership, or limited operational assistance. When conventional metrics assign full publication credit to every listed author, these heterogeneous roles can generate the same bibliometric reward as primary intellectual execution or senior scientific leadership.

The $\alpha$-index is therefore based on an authorship-integrity principle: strong credit should be attached to identifiable intellectual execution and accountable scientific leadership, while residual middle authorship should contribute limited credit unless a field-specific or paper-specific justification indicates otherwise. Acknowledgements should regain their role as a legitimate and visible place to recognize valuable non-authorial assistance, discussion, data access, technical support, and infrastructure provision. The index is not designed to label individual authors or papers as unethical. Rather, it is designed to correct an incentive structure that can make inflated authorship rational, professionally advantageous, and difficult to distinguish from necessary collaboration.

The senior-author penalty is the ethical core of the metric. If a senior author receives credit for leadership, then that author also carries responsibility for authorship discipline. Consequently, when the residual middle-author list expands, the senior-author credit decreases. This does not prove misconduct. It aims to provide a transparent signal that apparent leadership credit should be interpreted in light of byline expansion. In this sense, the $\alpha$-index is designed to make quality-over-quantity evaluation, authorship necessity, and accountability more visible, rather than to prove misconduct from byline order alone.

\section{Local $\alpha$-credit}

Consider a paper $p$ with $n_p$ authors ordered as
\[
A_{1,p}, A_{2,p}, \ldots, A_{n_p,p}.
\]
Each paper has total authorship value
\[
V_p=1.00.
\]
The local $\alpha$-credit of author $j$ on paper $p$ is
\[
\alpha_{j,p}\in[0,1],
\]
with the conservation constraint
\begin{equation}
\sum_{j=1}^{n_p}\alpha_{j,p}=1.
\label{eq:conservation}
\end{equation}

\subsection{Single-author and two-author papers}

If $n_p=1$, the sole author receives full credit:
\begin{equation}
\alpha_{1,p}=1.00.
\end{equation}

If $n_p=2$ and no equal-first-authorship declaration is present, the first author receives 0.67 and the second author receives 0.33:
\begin{equation}
(\alpha_{1,p},\alpha_{2,p})=(0.67,0.33).
\end{equation}

If $n_p=2$ and both authors are explicitly equal first authors, the allocation is
\begin{equation}
(\alpha_{1,p},\alpha_{2,p})=(0.50,0.50).
\end{equation}

\subsection{A baseline, non-penalized allocation for papers with more than two authors}

For $n_p>2$, let $k_F$ be the number of explicitly declared equal-first authors and let $k_L$ be the number of explicitly declared equal-last or equal-senior authors. If no equal-first or equal-last note is present, then $k_F=k_L=1$. Let
\begin{equation}
m=n_p-k_F-k_L
\end{equation}
be the number of residual middle authors.

The non-penalized baseline allocation uses three blocks:
\begin{equation}
F=0.67,\quad M=0.10,\quad L=0.23,
\label{eq:blocks}
\end{equation}
where $F$ denotes the first-author execution block, $M$ the residual middle-author block, and $L$ the senior-author leadership block. If $m>0$, the allocation is
\begin{equation}
\alpha_{j,p}=\frac{0.67}{k_F}\quad\text{for equal-first authors},
\end{equation}
\begin{equation}
\alpha_{j,p}=\frac{0.10}{m}\quad\text{for residual middle authors},
\end{equation}
\begin{equation}
\alpha_{j,p}=\frac{0.23}{k_L}\quad\text{for equal-last/equal-senior authors}.
\end{equation}
If $m=0$, the middle-author block is redistributed to the first-author execution block, giving
\begin{equation}
\alpha_{j,p}=\frac{0.77}{k_F}\quad\text{for equal-first authors},
\end{equation}
\begin{equation}
\alpha_{j,p}=\frac{0.23}{k_L}\quad\text{for equal-last/equal-senior authors}.
\end{equation}
This special case preserves Equation~\ref{eq:conservation} when there are no residual middle authors. Normatively, the default assigns the absent residual block to the execution side because, when no residual middle-author region exists, the framework treats the paper as maximally concentrated around the first/equal-first execution role. This is a default, not a necessity: a field that regards execution and senior leadership as equally strengthened by an absent middle region may instead adopt a symmetric or proportional redistribution within the parameterized family. The non-penalized allocation is retained as an analytical baseline: comparing it with the final penalized $\alpha$-index makes the specific effect of the senior-author responsibility penalty transparent.

\section{Global $\alpha$-index}

Let researcher $i$ have $M_i$ publications. In this manuscript, the $\alpha$-index always means the penalized cumulative measure defined below; the unpenalized allocation is retained solely as a transparent baseline. The final $\alpha$-index is the cumulative sum of the penalized local $\alpha$-credits:
\begin{equation}
\alpha_i^{G}=\sum_{p=1}^{M_i}\alpha^{P}_{i,p}.
\label{eq:global_alpha}
\end{equation}
Thus, $\alpha_i^{G}$ measures accumulated contribution-weighted authorship credit under the penalized allocation. If $M_i=1$, then
\begin{equation}
\alpha_i^{G}=\alpha^{P}_{i,1},
\end{equation}
so the global $\alpha$-index equals the local penalized $\alpha$-credit for a single-paper author.

The $\alpha$-index is deliberately cumulative rather than averaged. It therefore distinguishes an author with one high-credit paper from an author who has accumulated substantial contribution-weighted credit across many papers. It is not a citation-impact metric and should be interpreted alongside publication count, citation-based indicators, peer review, and contribution statements.

\section{The $\alpha$-index: penalized senior-author responsibility}
\label{sec:penalized}

The baseline allocation is useful as a local and global authorship-credit foundation. However, the main proposed metric of this paper is the $\alpha$-index, denoted by $\alpha^P$ when the penalty needs to be made explicit. The final penalized allocation formalizes the idea that last or senior authors are not only leaders of the project but also custodians of authorship discipline. If a paper has many middle authors, the senior author may still have led the project, but the metric should not reward the senior author with the same fixed leadership credit regardless of how much the byline has expanded.

For $n_p\leq2$, the penalized $\alpha$-index equals the corresponding local $\alpha$-credit because no senior-author dilution is applied. For $n_p>2$, using the notation above, let
\[
m=n_p-k_F-k_L
\]
be the number of residual middle authors.

If $m=0$, there are no residual middle authors and no senior-author dilution is applied. Under the default redistribution rule, the absent middle block is assigned to the first-author execution side rather than to senior authorship. The allocation is therefore the same as the unpenalized local baseline:
\begin{equation}
\alpha^{P}_{j,p}=\frac{0.77}{k_F}\quad\text{for equal-first authors},
\end{equation}
\begin{equation}
\alpha^{P}_{j,p}=\frac{0.23}{k_L}\quad\text{for equal-last/equal-senior authors}.
\end{equation}
The $m=0$ redistribution is deliberately asymmetric: it reflects the default stance that execution credit should be maximized when no residual middle-authorship region exists. It is not asserted as universally correct; a symmetric or proportional redistribution between the execution and senior blocks is an admissible field-specific variant of the same conserved family.

If $m>0$, the first-author execution block remains fixed:
\begin{equation}
F^{P}=0.67.
\end{equation}
The total senior-author block is then diluted according to the number of residual middle authors:
\begin{equation}
L^{P}(m)=\frac{0.23}{m}.
\label{eq:lp}
\end{equation}
The remaining credit is assigned to the residual middle-author pool:
\begin{equation}
M^{P}(m)=1-0.67-L^{P}(m)=0.33-\frac{0.23}{m}.
\label{eq:mp}
\end{equation}
Thus the paper-level penalized allocation is
\begin{equation}
\alpha^{P}_{j,p}=\frac{0.67}{k_F}\quad\text{for equal-first authors},
\end{equation}
\begin{equation}
\alpha^{P}_{j,p}=\frac{M^{P}(m)}{m}=\frac{0.33-0.23/m}{m}\quad\text{for residual middle authors},
\end{equation}
\begin{equation}
\alpha^{P}_{j,p}=\frac{L^{P}(m)}{k_L}=\frac{0.23}{m k_L}\quad\text{for equal-last/equal-senior authors}.
\end{equation}
The conservation constraint is preserved:
\begin{equation}
\sum_{j=1}^{n_p}\alpha^{P}_{j,p}=1.
\end{equation}

A direct consequence of Equation~\ref{eq:mp} is that the collective residual middle-author pool increases as the senior-author block decreases:
\begin{equation}
M^{P}(m)=1-F-\frac{L}{m}.
\end{equation}
Under the default values, \(M^{P}(m)=0.33-0.23/m\), so \(M^{P}(1)=0.10\), \(M^{P}(2)=0.215\), and \(M^{P}(m)\to0.33\) as \(m\to\infty\). This behavior is intentional in the strict default allocation: the released credit is not transferred to the protected first-author block; instead, it remains inside the conserved paper credit and is redistributed to the residual byline. The penalty therefore acts primarily on senior-author leadership credit, not on the existence of middle authors themselves. In applications where an asymptotic middle pool of \(1-F\) is judged too generous or too severe, a softer penalty function, a capped residual pool, or field-specific calibration should be used.

The global $\alpha$-index is obtained by summing these penalized local credits across an author's publication record, as defined in Equation~\ref{eq:global_alpha}. In the remainder of the manuscript, $\alpha_i^{G}$ denotes the official cumulative global $\alpha$-index under the default penalized allocation unless otherwise specified. This definition separates accumulated contribution-weighted credit from raw publication count without introducing an average-role statistic.

\subsection{Interpretation}

Equation~\ref{eq:lp} is the norm-expressing and incentive-structuring design choice in the metric. It is intended to make the senior-author cost of residual middle-author expansion visible; whether it changes authorship behavior is an empirical question for future work. When there is one middle author, the senior block remains 0.23. When there are two residual middle authors, it falls to 0.115. When there are ten residual middle authors, it falls to 0.023. Thus, the first-author execution block is protected, but senior-author credit is progressively diluted as the middle-author list grows.

This does not imply that all middle authors are illegitimate. It means that the index distinguishes authorship from acknowledgement-level contribution. Essential intellectual contributors can be recognized as equal-first authors, equal-senior authors, or substantive coauthors whose credit is visible within a conserved allocation. Routine technical help, data access, recruitment support, infrastructure provision, administrative support, and general collaboration can still be acknowledged without necessarily generating full bibliometric authorship credit.

\subsection{Alternative penalty functions and large-team science}
\label{sec:penalty_variants}

The strict hyperbolic penalty in Equation~\ref{eq:lp} intentionally makes senior-author credit fall quickly as the residual middle-author list expands. This is the strongest authorship-discipline version of the $\alpha$-index. However, it may be too severe for legitimate large-team science, including multicenter clinical trials, genomics consortia, high-energy physics collaborations, and infrastructure-heavy biomedical studies. For this reason, implementations should report the default strict value but may also flag papers for field-specific interpretation or compute softer variants.

Three representative senior-author penalty functions are:
\begin{align}
\text{strict hyperbolic:}\quad & L^{P}_{\mathrm{hyp}}(m)=\frac{L}{m}, \\
\text{square-root:}\quad & L^{P}_{\sqrt{}}(m)=\frac{L}{\sqrt{m}}, \\
\text{logarithmic:}\quad & L^{P}_{\log}(m)=\frac{L}{\log_2(m+1)}.
\end{align}
All three preserve $L^P(1)=L$ and reduce senior credit as $m$ increases, but they differ in severity. The strict hyperbolic version is the official default $\alpha$-index because it most directly encodes the authorship-discipline principle. The square-root and logarithmic variants are not proposed as replacements here; they are calibration candidates for fields where large residual author lists can be structurally necessary.

A practical implementation should also include a large-team flag. When $m$ exceeds a prespecified field-specific threshold, the paper can be annotated as ``large-team or consortium science.'' The threshold should be selected from the empirical author-list distribution of the field under study (for example, the 90th or 95th percentile of residual middle-author counts estimated in a documented bibliographic calibration sample), rather than treated as a universal cut point. Thus, the threshold is a calibration output of a field-specific bibliographic analysis, not an assumed property of the framework. This makes the threshold reproducible without pretending that one number suits every discipline. The index can still be computed, but evaluation should not rely on the numerical value alone. This avoids the false interpretation that the metric is adjudicating misconduct in structurally collaborative research. It instead signals that ordinary byline-order credit is insufficient and that contribution statements, consortium policies, and field norms must be consulted.

\subsection{Potential unintended incentives}

A penalized metric can create incentives. One possible concern is acknowledgement substitution: a senior author might be tempted to move a legitimate middle contributor into the acknowledgements to preserve senior-author credit. This would violate the purpose of the index. The $\alpha$-index is not intended to discourage proper authorship. It is intended to discourage unnecessary authorship. Legitimate contributors who satisfy authorship criteria should remain authors, and essential intellectual contributors may warrant equal-first, equal-senior, or clearly documented contributor roles. The correct use of the index is therefore conditional on ethical authorship practice and should be combined with ICMJE/COPE guidance and CRediT-style contribution statements.

\section{Algorithms}

\begin{algorithm}[H]
\caption{Parameterized penalized local $\alpha$-allocation. The default $\alpha$-index uses $F_2=0.67$, $F=0.67$, $M=0.10$, $L=0.23$, and $g(m)=1/m$.}
\label{alg:penalized}
\begin{algorithmic}[1]
\Require Number of authors $n$, number of equal first authors $k_F$, number of equal senior authors $k_L$, two-author block $F_2$, blocks $F,M,L$ with $F+M+L=1$, penalty function $g(m)$ with $g(1)=1$
\Ensure Vector $\boldsymbol{\alpha}^{P}=(\alpha^{P}_1,\ldots,\alpha^{P}_n)$ with $\sum_j\alpha^{P}_j=1$
\If{$n=1$}
  \State \Return $(1.00)$
\EndIf
\If{$n=2$ and $k_F=2$}
  \State \Return $(0.50,0.50)$
\EndIf
\If{$n=2$ and $k_F=1$}
  \State \Return $(F_2,1-F_2)$ \Comment{Default: $F_2=0.67$}
\EndIf
\State $m\gets n-k_F-k_L$
\If{$m=0$}
  \State Assign $(F+M)/k_F$ to each equal-first author \Comment{Default: maximize execution credit when no residual middle region exists}
  \State Assign $L/k_L$ to each equal-senior author
\Else
  \State $L^{P}\gets L\,g(m)$
  \State $M^{P}\gets 1-F-L^{P}$
  \State Assign $F/k_F$ to each equal-first author
  \State Assign $M^{P}/m$ to each residual middle author
  \State Assign $L^{P}/k_L$ to each equal-senior author
\EndIf
\State \Return $\boldsymbol{\alpha}^{P}$
\end{algorithmic}
\end{algorithm}

\section{Synthetic examples}

\subsection{Local allocations under the penalized index}

Table~\ref{tab:pen_examples} shows how the $\alpha$-index behaves across representative authorship structures.

\begin{table}[ht]
\centering
\caption{Penalized $\alpha$-index allocations across representative authorship configurations. EF denotes equal-first authors and ES denotes equal-senior/equal-last authors.}
\label{tab:pen_examples}
\small
\resizebox{\textwidth}{!}{%
\begin{tabular}{@{}lllll@{}}
\toprule
Case & Configuration & EF & ES & Penalized allocation by author order \\
\midrule
1 & One author & 1 & 0 & $(1.000)$ \\
2 & Two authors, no equal first & 1 & 1 & $(0.670,0.330)$ \\
3 & Two authors, equal first & 2 & 0 & $(0.500,0.500)$ \\
4 & Three authors & 1 & 1 & $(0.670,0.100,0.230)$ \\
5 & Four authors & 1 & 1 & $(0.670,0.1075,0.1075,0.115)$ \\
6 & Six authors & 1 & 1 & $(0.670,0.0681,0.0681,0.0681,0.0681,0.0575)$ \\
7 & Five authors, two equal first & 2 & 1 & $(0.335,0.335,0.1075,0.1075,0.115)$ \\
8 & Three authors, two equal first, no middle & 2 & 1 & $(0.385,0.385,0.230)$ \\
9 & Six authors, two equal senior & 1 & 2 & $(0.670,0.0844,0.0844,0.0844,0.0383,0.0383)$ \\
\bottomrule
\end{tabular}%
}
\end{table}

The transition from Case 4 to Case 6 illustrates the key property. The first author remains at 0.67, but the last author falls from 0.23 to 0.0575 as the number of middle authors increases from one to four. The released credit is not lost; it is redistributed within the conserved paper value. The metric therefore discourages senior authors from expanding middle authorship while still recognizing that each additional middle author may have some contribution.

\subsection{Comparison with standard counting schemes}

Table~\ref{tab:method_comparison} compares full counting, equal fractional counting, harmonic counting, the non-penalized baseline allocation, and the $\alpha$-index for a six-author paper without equal-contribution notes.

\begin{table}[ht]
\centering
\caption{Comparison of author-credit allocation methods for a six-author paper without equal first or equal senior authorship. Harmonic credit is calculated as $(1/r)/H_n$, where $r$ is author rank and $H_n$ is the $n$th harmonic number.}
\label{tab:method_comparison}
\small
\resizebox{\textwidth}{!}{%
\begin{tabular}{@{}lcccccc@{}}
\toprule
Method & Author 1 & Author 2 & Author 3 & Author 4 & Author 5 & Author 6 \\
\midrule
Full counting & 1.000 & 1.000 & 1.000 & 1.000 & 1.000 & 1.000 \\
Equal fractional & 0.167 & 0.167 & 0.167 & 0.167 & 0.167 & 0.167 \\
Harmonic & 0.408 & 0.204 & 0.136 & 0.102 & 0.082 & 0.068 \\
Non-penalized baseline & 0.670 & 0.025 & 0.025 & 0.025 & 0.025 & 0.230 \\
$\alpha$-index (penalized) & 0.670 & 0.068 & 0.068 & 0.068 & 0.068 & 0.058 \\
\bottomrule
\end{tabular}%
}
\end{table}

Full counting violates conservation because one paper creates six author credits. Equal fractional counting conserves credit but ignores authorship role. Harmonic counting rewards earlier author positions but makes the last author the smallest-credit position. The non-penalized baseline encodes first-author and last-author conventions. The $\alpha$-index retains the first-author execution block while reducing last-author credit when the middle-author list expands.

\subsection{Synthetic author profiles and sensitivity analysis}

Table~\ref{tab:author_profiles} shows how different author profiles separate under full counting and under the $\alpha$-index. The table deliberately excludes h-index values because the h-index depends on citations, not byline position. The $\alpha$-index should be read as an authorship-integrity indicator, not as a citation-impact metric.

\begin{table}[ht]
\centering
\caption{Synthetic author profiles illustrating how the $\alpha$-index distinguishes authorship patterns that may look similar under full counting.}
\label{tab:author_profiles}
\scriptsize
\resizebox{\textwidth}{!}{%
\begin{tabular}{@{}p{2.8cm}p{4.7cm}rrrrr@{}}
\toprule
Profile & Synthetic publication pattern & Papers & Full count & Equal fractional total & Harmonic total & $\alpha^{G}$ \\
\midrule
Execution-focused author & 10 first-author papers, each with 6 authors & 10 & 10.00 & 1.67 & 4.08 & 6.70 \\
Senior with compact teams & 10 last-author papers, each with 3 authors & 10 & 10.00 & 3.33 & 1.82 & 2.30 \\
Senior with expanded teams & 10 last-author papers, each with 10 authors & 10 & 10.00 & 1.00 & 0.34 & 0.288 \\
Middle-list collaborator & 10 middle-author papers, each with 6 authors & 10 & 10.00 & 1.67 & 1.02 & 0.681 \\
Independent author & 4 single-author papers & 4 & 4.00 & 4.00 & 4.00 & 4.00 \\
Balanced author & 5 first-author and 5 compact last-author papers & 10 & 10.00 & 2.50 & 2.95 & 4.50 \\
\bottomrule
\end{tabular}%
}
\end{table}

The key contrast is between the two senior profiles. Under full counting, they are identical. Under the $\alpha$-index, the senior author with compact teams receives substantially higher global credit than the senior author who repeatedly appears last on very long author lists. This is the intended behavior: leadership credit is preserved when the author list is disciplined and diluted when middle authorship expands.

Table~\ref{tab:sensitivity} provides a simple sensitivity analysis over plausible block choices. The profiles are intentionally stylized, so the table should not be interpreted as validation. Its purpose is to show whether the main ranking signal depends entirely on the default constants. Across the displayed settings, execution-focused and independent authors remain high-credit profiles, compact senior leadership remains above expanded senior leadership, and repeated expanded senior authorship remains strongly penalized.

\begin{table}[ht]
\centering
\caption{Sensitivity analysis of the cumulative global $\alpha$-index $\alpha^{G}$ under alternative block settings. Each triplet is $(F,M,L)$ with $F+M+L=1$ and the same strict penalty $L/m$.}
\label{tab:sensitivity}
\scriptsize
\resizebox{\textwidth}{!}{%
\begin{tabular}{@{}p{3.0cm}rrrr@{}}
\toprule
Profile & $(0.60,0.15,0.25)$ & $(0.67,0.10,0.23)$ default & $(0.70,0.10,0.20)$ & $(0.75,0.05,0.20)$ \\
\midrule
Execution-focused author & 6.00 & 6.70 & 7.00 & 7.50 \\
Senior with compact teams & 2.50 & 2.30 & 2.00 & 2.00 \\
Senior with expanded teams & 0.313 & 0.288 & 0.250 & 0.250 \\
Middle-list collaborator & 0.844 & 0.681 & 0.625 & 0.500 \\
Independent author & 4.00 & 4.00 & 4.00 & 4.00 \\
Balanced author & 4.25 & 4.50 & 4.50 & 4.75 \\
\bottomrule
\end{tabular}%
}
\end{table}

The sensitivity analysis also highlights the need for empirical calibration. The default constants influence absolute values, but the qualitative signal targeted by the index is stable in these examples: compact execution and disciplined leadership receive more credit than repeated long-list authorship. Future empirical work should test this robustness in real bibliographic datasets.

Table~\ref{tab:penalty_sensitivity} examines the second calibration axis: the penalty function $g(m)$. The strict hyperbolic function produces the largest separation between compact and expanded senior authorship. Square-root and logarithmic penalties retain the same direction of effect but reduce the severity for large-team settings. This comparison is not empirical validation; it shows how implementation choices alter the strength of the authorship-discipline signal.

\begin{table}[ht]
\centering
\caption{Penalty-function sensitivity analysis under the default block settings $(F_2,F,M,L)=(0.67,0.67,0.10,0.23)$. Values are cumulative global $\alpha$-indices $\alpha_i^{G}$ under the penalized allocation for the same stylized author profiles.}
\label{tab:penalty_sensitivity}
\scriptsize
\resizebox{\textwidth}{!}{%
\begin{tabular}{@{}p{3.4cm}rrr@{}}
\toprule
Profile & Strict $g(m)=1/m$ & Square-root $g(m)=1/\sqrt{m}$ & Logarithmic $g(m)=1/\log_2(m+1)$ \\
\midrule
Execution-focused author & 6.700 & 6.700 & 6.700 \\
Senior with compact teams & 2.300 & 2.300 & 2.300 \\
Senior with expanded teams & 0.288 & 0.813 & 0.726 \\
Middle-list collaborator & 0.681 & 0.537 & 0.577 \\
Independent author & 4.000 & 4.000 & 4.000 \\
Balanced author & 4.500 & 4.500 & 4.500 \\
\bottomrule
\end{tabular}%
}
\end{table}

\section{Illustrative real-publication case studies}
\label{sec:real_cases}

This section applies the penalized local $\alpha$-index to selected real publications from biomedical research and computer science. These calculations are byline-structure demonstrations, not judgments about the named researchers' true contributions. Complete empirical evaluation would require author disambiguation, equal-contribution metadata extraction, corresponding-author notes, contribution statements, and field-specific validation.

The examples include the ResNet paper by He et al. \citep{He2016}, the Transformer paper by Vaswani et al. \citep{Vaswani2017}, the BERT paper by Devlin et al. \citep{Devlin2019}, the diabetic-retinopathy deep-learning paper by Gulshan et al. \citep{Gulshan2016}, the dermatologist-level skin-cancer paper by Esteva et al. \citep{Esteva2017}, the AlphaFold paper by Jumper et al. \citep{Jumper2021}, and a Nature Medicine diabetes-care AI paper with a long author list and three equal-contributing first authors \citep{Li2024NatureMedicine}.

\begin{table}[ht]
\centering
\caption{Illustrative $\alpha$-index calculations on selected real publications. EF denotes modeled equal-first authors and ES denotes modeled equal-senior authors. ``Middle each'' is the value for each residual middle author.}
\label{tab:real_cases}
\scriptsize
\resizebox{\textwidth}{!}{%
\begin{tabular}{@{}p{3.2cm}p{2.0cm}rrrrrrr@{}}
\toprule
Publication & Field & Authors $n$ & EF & ES & First/equal-first each & Middle each & Senior each & Sum \\
\midrule
He et al., ResNet \citep{He2016} & Computer vision & 4 & 1 & 1 & 0.670 & 0.1075 & 0.1150 & 1.000 \\
Vaswani et al., Transformer \citep{Vaswani2017} & NLP / CS & 8 & 1 & 1 & 0.670 & 0.0486 & 0.0383 & 1.000 \\
Devlin et al., BERT \citep{Devlin2019} & NLP / CS & 4 & 1 & 1 & 0.670 & 0.1075 & 0.1150 & 1.000 \\
Gulshan et al., diabetic retinopathy \citep{Gulshan2016} & Biomedical AI & 15 & 1 & 1 & 0.670 & 0.0240 & 0.0177 & 1.000 \\
Esteva et al., skin cancer \citep{Esteva2017} & Biomedical AI & 7 & 2 & 1 & 0.335 & 0.0681 & 0.0575 & 1.000 \\
Jumper et al., AlphaFold, byline-only model \citep{Jumper2021} & Computational biology & 34 & 1 & 1 & 0.670 & 0.0101 & 0.0072 & 1.000 \\
Li et al., diabetes-care AI \citep{Li2024NatureMedicine} & Biomedical AI & 93 & 3 & 1 & 0.223 & 0.0037 & 0.0026 & 1.000 \\
\bottomrule
\end{tabular}%
}
\end{table}

These examples demonstrate the mechanical behavior of the penalized index. A four-author computer-vision paper gives the senior author 0.115, while a 93-author biomedical AI paper gives the single senior author approximately 0.0026 under the modeled byline structure. The first-author or equal-first-author execution block remains protected, but senior-author credit becomes small when the residual middle-author list becomes large. This should not be read as a judgment about those specific papers or authors. It shows why large-team cases require contribution metadata, field norms, and possibly softer or capped penalty variants.

The face-validity interpretation is deliberately modest. Compact computer-science bylines such as the ResNet and BERT examples produce a visible senior-author value, while very large biomedical or computational-biology bylines produce very small strict senior-author values. This is consistent with the intended behavior of the strict index: a byline-only model should not assign the same leadership credit to a single senior position in a 93-author paper as in a compact three- or four-author paper. At the same time, the AlphaFold and long biomedical-AI examples illustrate the limits of byline-only evaluation. Their low senior values are best read as flags that contribution metadata and field-specific variants are required, not as claims about the actual intellectual contribution of any named author.

Some papers require careful curation. For example, papers with broad equal-contribution notes, consortium authorship, or overlapping equal-first and equal-senior roles may not be adequately represented by simple byline parsing. Such cases should be flagged rather than forced into an automated calculation.

\section{Relationship to existing indicators}

The $\alpha$-index framework is closest to authorship-credit allocation methods, not to citation-impact metrics. Table~\ref{tab:related_indices} summarizes the conceptual relationship.

\begin{table}[ht]
\centering
\caption{Conceptual comparison of related indicators.}
\label{tab:related_indices}
\scriptsize
\begin{tabular}{@{}p{2.7cm}p{2.7cm}p{2.2cm}p{5.8cm}@{}}
\toprule
Indicator or family & Main object & Direct competitor? & Key distinction from the $\alpha$-index \\
\midrule
h-index \citep{Hirsch2005} & Citation impact and productivity & No & Full coauthor credit; no local authorship-credit coefficient. \\
g-index / e-index \citep{Egghe2006,Zhang2009} & Citation-impact refinement & No & Citation-focused; does not allocate publication credit by authorship role. \\
Full counting & Publication or citation count & Yes & Violates conservation because one paper creates $n$ author credits. \\
Equal fractional counting & Conserved publication credit & Yes & Conserves credit but treats first, middle, and last authors equally. \\
Rank-based / harmonic counting \citep{Hagen2008,Hagen2010} & Ordered author credit & Yes & Rewards author rank but does not preserve elevated senior-author credit under compact authorship or penalize senior authorship under expansion in the same way. \\
$h_I$, $h_P$, $h_m$ \citep{Batista2006,Wan2007,Schreiber2008} & Corrected h-type impact & Partial & Modify h-type indicators; the $\alpha$-index can be computed without citations. \\
$h_{\alpha}$ \citep{Hirsch2019alpha} & Leadership by highest-h coauthor & Partial & Assigns leadership based on coauthors' h-indices; the $\alpha$-index uses local byline structure and conserved credit. \\
Relative/weighted rankings \citep{Vavrycuk2018} & Authorship inflation correction & Yes & Related motivation, but different allocation rule, local/global structure, and senior-author penalty. \\
$\alpha$-index (penalized allocation) $\alpha^P$ & Authorship integrity & Proposed method & Conserved, first-author-protective, last-author-aware, citation-independent, and penalizes senior credit when middle authorship expands. \\
\bottomrule
\end{tabular}
\end{table}

The most important distinction is normative and incentive-oriented. Many existing methods correct counting bias after publication. The $\alpha$-index is designed to make the senior-author cost of residual middle-author expansion explicit, thereby making authorship discipline visible before and after publication.

The distinction from Hirsch's $h_{\alpha}$ is especially important. The $h_{\alpha}$ framework identifies an ``alpha author'' of a paper using the highest h-index among coauthors, whereas the $\alpha$-index assigns conserved local credit from the byline structure itself. In a stylized four-author paper with authorship order A--B--C--D, suppose D has the highest h-index. An $h_{\alpha}$-style interpretation attributes alpha status to D, independent of the conserved distribution of work. Under the default $\alpha$-allocation, the same paper gives A a local execution credit of $0.67$, B and C $0.1075$ each, and D $0.115$. The two metrics therefore answer different questions: $h_{\alpha}$ asks which coauthor has the strongest prior citation record, whereas the $\alpha$-index asks how one unit of authorship credit should be distributed for this paper under a position-weighted authorship-integrity rule.

\section{Interpretation and reporting recommendations}

The cumulative global $\alpha$-index $\alpha_i^{G}$ is an accumulated authorship-integrity credit, not a comprehensive impact metric. It measures the total conserved authorship credit an author accumulates under the penalized allocation. It should not be interpreted as citation influence, methodological quality, clinical importance, societal value, or scientific correctness.

For evaluation, the preferred reporting unit is
\begin{equation}
\left(M_i,h_i,\alpha_i^{G}\right),
\end{equation}
where $M_i$ is publication count and $h_i$ is the conventional h-index or another field-normalized impact measure. The value $\alpha_i^{G}$ measures total contribution-weighted productivity under the authorship-discipline penalty. When possible, this summary should also report the number and percentage of sole-author, first-author, equal-first-author, last-author, equal-last-author, middle-author, and consortium papers.

The $\alpha$-index is complementary to citation-based metrics. Two authors with identical $\alpha$-indices may have very different citation impact. Conversely, two authors with identical h-indices may have very different authorship structures. The proposed metric addresses only the latter dimension.

\section{Applications}

The $\alpha$-index could be used in researcher self-assessment, promotion dossiers, grant review, bibliometric dashboards, institutional reporting, and research-integrity audits, but only as a complement to qualitative assessment. It may be especially useful when two candidates have similar publication counts or citation metrics but very different authorship profiles. It can also encourage journals, institutions, and research groups to restore the distinction between authorship and acknowledgement.

A scalable implementation would require author-order parsing, equal-contribution metadata extraction, author disambiguation, and field-specific rules for consortium papers. The most realistic near-term use is therefore not automated universal ranking, but transparent profile reporting for curated publication lists. A future database could compute $\alpha$-profiles from PubMed, DBLP, OpenAlex, ORCID, Crossref, Semantic Scholar, or institutional publication repositories, while flagging records with missing or ambiguous equal-contribution metadata.

The metric should be interpreted within field norms. It is not suitable without adaptation for alphabetical authorship fields, formal high-energy-physics-style collaborations, or settings where author order is not meaningful. For large-team science, the numerical value should be accompanied by a large-team flag and, when available, contribution-statement evidence.

\section{Limitations}

The first and most important limitation is that the default weights are normative. The values 0.67, 0.10, and 0.23 express a particular stance: primary execution deserves dominant credit, residual middle authorship should receive limited credit, and senior leadership deserves credit but should be diluted when middle authorship expands. These values are transparent defaults, not empirically established constants. The framework should be calibrated and stress-tested in different disciplines.

Second, the strict penalty $L/m$ is intentionally strong. It can appear too severe for legitimate large-scale science. This manuscript therefore presents square-root and logarithmic alternatives and recommends large-team flags. The strict version should be interpreted as the authorship-discipline default, not as an accusation of misconduct.

Third, author order is an imperfect proxy for contribution. Some middle authors make essential conceptual, methodological, clinical, computational, or experimental contributions. Some first authors may not be the true intellectual drivers of a paper. Some last authors may be ceremonial. No position-based index can fully solve these problems. Contribution statements, acknowledgements, peer review, and expert judgment remain necessary.

Fourth, equal-contribution metadata are inconsistently indexed in bibliographic databases. Equal-first and equal-senior declarations may appear in PDF footnotes, publisher metadata, PubMed notes, institutional repositories, or not at all. Automated implementation will therefore produce errors unless it includes confidence flags and manual curation options.

Fifth, the index may create unintended incentives. In particular, if misused, a senior author might try to preserve credit by moving legitimate contributors to acknowledgements. This would be contrary to the purpose of the index and contrary to authorship ethics. The metric should be used to discourage unnecessary authorship, not to deny proper authorship to genuine contributors.

Finally, the index does not measure citation impact, methodological rigor, reproducibility, scientific correctness, clinical utility, or societal value. It is an authorship-structure metric. It should be combined with citation-based indicators, peer review, contribution statements, and qualitative assessment whenever used in research evaluation.

\section{Validation roadmap and future work}

The $\alpha$-index should next be evaluated empirically. A minimal validation roadmap includes four components. First, sensitivity analyses should be performed across plausible $(F,M,L)$ triplets and penalty functions. Second, the metric should be compared with CRediT statements in journals where contributor roles are available, for example by mapping conceptualization, methodology, formal analysis, investigation, data curation, software, supervision, and writing roles to independently rated contribution profiles. Third, author surveys and expert panels should be used to test whether the index aligns with perceived intellectual contribution and leadership. Fourth, large-scale bibliographic studies should examine distributions of $\alpha_i^{G}$ across fields, career stages, and collaboration cultures.

A useful empirical study could sample 10,000--50,000 papers from a well-defined field such as biomedical artificial intelligence, epidemiology, computer vision, or natural language processing. For each author, one could compute publication count, full counting, equal fractional counting, harmonic credit, h-index or field-normalized citation impact, and $\alpha$-profiles. The scientific value of the metric would then be assessed by where rankings converge and diverge, and whether divergences correspond to meaningful differences in authorship role.

Future work should develop an open-source implementation with explicit handling of equal-first authorship, equal-senior authorship, consortium flags, missing metadata, and alternative penalty functions. A public database could eventually allow authors to inspect and correct their own $\alpha$-profiles, provided that author disambiguation and metadata provenance are transparent.

The index could also be integrated with citation impact through a field-normalized extension,
\begin{equation}
\alpha\text{-citation}_{i,p}^{P}=\alpha^{P}_{i,p}\times C_p^{\ast},
\end{equation}
where $C_p^{\ast}$ is a field-normalized citation measure. This extension would preserve the distinction between authorship structure and impact while allowing both dimensions to be reported jointly.

\section{Conclusion}

We introduced the $\alpha$-index, denoted by the Greek lowercase letter \asymbol, as a conserved, penalized authorship-integrity framework. The local allocation assigns one unit of credit across a publication's authors, and the cumulative global $\alpha$-index summarizes an author's accumulated contribution-weighted authorship credit. The central innovation is the senior-author responsibility penalty, which reduces senior-author credit as the residual middle-author list expands.

The purpose of the index is not to declare true contribution from byline order alone. Rather, it is to make authorship structure visible and to formalize a normative principle: identifiable intellectual execution and disciplined scientific leadership should receive more credit than accumulated authorship volume. The paper proposes the default $\alpha$-index as a transparent starting point, while explicitly recognizing the need for sensitivity analysis, field-specific calibration, contribution-statement validation, and careful treatment of large-team science.

The $\alpha$-index does not replace peer review, contribution statements, acknowledgements, or citation-based metrics. Its role is complementary: it aims to provide a reproducible way to report authorship integrity and to reopen the distinction between authorship-level contribution and acknowledgement-level support.

\section*{Data availability}
No private or newly collected empirical data are analyzed. The manuscript includes synthetic examples and illustrative calculations using publicly available byline metadata from selected publications. These examples are intended as methodological demonstrations rather than validation studies or complete evaluations of named authors.

\section*{Code availability}
No software accompanies this preprint. A documented reference implementation is being prepared for release in a public repository. The present manuscript contains the complete mathematical specification, algorithm, and worked calculations needed to reproduce the allocation rules.

\section*{Conflict of interest}
No financial conflict of interest is declared.

\section*{Funding}
No external funding was received for this work.

\section*{Author contributions}
Athanasios Angelakis conceptualized the $\alpha$-index, developed the local and cumulative penalized metric definitions, prepared the examples, and wrote the manuscript.


\begin{thebibliography}{99}

\bibitem[Hirsch(2005)]{Hirsch2005}
Hirsch, J. E. (2005).
An index to quantify an individual's scientific research output.
\emph{Proceedings of the National Academy of Sciences}, 102(46), 16569--16572.
\href{https://doi.org/10.1073/pnas.0507655102}{https://doi.org/10.1073/pnas.0507655102}.

\bibitem[ICMJE(2024)]{ICMJE2024}
International Committee of Medical Journal Editors. (2024).
Recommendations for the Conduct, Reporting, Editing, and Publication of Scholarly Work in Medical Journals: Defining the Role of Authors and Contributors.
\url{https://www.icmje.org/recommendations/browse/roles-and-responsibilities/defining-the-role-of-authors-and-contributors.html}.

\bibitem[Brand et~al.(2015)]{Brand2015}
Brand, A., Allen, L., Altman, M., Hlava, M., and Scott, J. (2015).
Beyond authorship: attribution, contribution, collaboration, and credit.
\emph{Learned Publishing}, 28(2), 151--155.
\href{https://doi.org/10.1087/20150211}{https://doi.org/10.1087/20150211}.

\bibitem[Batista et~al.(2006)]{Batista2006}
Batista, P. D., Campiteli, M. G., Kinouchi, O., and Martinez, A. S. (2006).
Is it possible to compare researchers with different scientific interests?
\emph{Scientometrics}, 68(1), 179--189.
\href{https://doi.org/10.1007/s11192-006-0090-4}{https://doi.org/10.1007/s11192-006-0090-4}.

\bibitem[Wan et~al.(2007)]{Wan2007}
Wan, J.-K., Hua, P.-H., and Rousseau, R. (2007).
The pure h-index: calculating an author's h-index by taking co-authors into account.
\emph{COLLNET Journal of Scientometrics and Information Management}, 1(2), 1--5.
\href{https://doi.org/10.1080/09737766.2007.10700824}{https://doi.org/10.1080/09737766.2007.10700824}.

\bibitem[Schreiber(2008)]{Schreiber2008}
Schreiber, M. (2008).
A modification of the h-index: The hm-index accounts for multi-authored manuscripts.
\emph{Journal of Informetrics}, 2(3), 211--216.
\href{https://doi.org/10.1016/j.joi.2008.05.001}{https://doi.org/10.1016/j.joi.2008.05.001}.

\bibitem[Hagen(2008)]{Hagen2008}
Hagen, N. T. (2008).
Harmonic allocation of authorship credit: source-level correction of bibliometric bias assures accurate publication and citation analysis.
\emph{PLOS ONE}, 3(12), e4021.
\href{https://doi.org/10.1371/journal.pone.0004021}{https://doi.org/10.1371/journal.pone.0004021}.

\bibitem[Hagen(2010)]{Hagen2010}
Hagen, N. T. (2010).
Harmonic publication and citation counting: sharing authorship credit equitably--not equally, geometrically or arithmetically.
\emph{Scientometrics}, 84(3), 785--793.
\href{https://doi.org/10.1007/s11192-009-0129-4}{https://doi.org/10.1007/s11192-009-0129-4}.

\bibitem[Vavry\v{c}uk(2018)]{Vavrycuk2018}
Vavry\v{c}uk, V. (2018).
Fair ranking of researchers and research teams.
\emph{PLOS ONE}, 13(4), e0195509.
\href{https://doi.org/10.1371/journal.pone.0195509}{https://doi.org/10.1371/journal.pone.0195509}.

\bibitem[Sundling(2023)]{Sundling2023}
Sundling, P. (2023).
Author contributions and allocation of authorship credit: testing the validity of different counting methods in chemical biology.
\emph{Scientometrics}, 128, 2951--2979.
\href{https://doi.org/10.1007/s11192-023-04680-y}{https://doi.org/10.1007/s11192-023-04680-y}.

\bibitem[COPE Council(2019)]{COPE2019}
COPE Council. (2019).
COPE Discussion Document: Authorship.
\url{https://publicationethics.org/resources/discussion-documents/authorship}.

\bibitem[COPE(2021)]{COPEGift}
COPE Council. (2021).
Gift authorship.
\url{https://publicationethics.org/news/gift-authorship}.

\bibitem[Flanagin et~al.(1998)]{Flanagin1998}
Flanagin, A., Carey, L. A., Fontanarosa, P. B., Phillips, S. G., Pace, B. P., Lundberg, G. D., and Rennie, D. (1998).
Prevalence of articles with honorary authors and ghost authors in peer-reviewed medical journals.
\emph{JAMA}, 280(3), 222--224.
\href{https://doi.org/10.1001/jama.280.3.222}{https://doi.org/10.1001/jama.280.3.222}.

\bibitem[Wislar et~al.(2011)]{Wislar2011}
Wislar, J. S., Flanagin, A., Fontanarosa, P. B., and DeAngelis, C. D. (2011).
Honorary and ghost authorship in high impact biomedical journals: a cross sectional survey.
\emph{BMJ}, 343, d6128.
\href{https://doi.org/10.1136/bmj.d6128}{https://doi.org/10.1136/bmj.d6128}.

\bibitem[He et~al.(2016)]{He2016}
He, K., Zhang, X., Ren, S., and Sun, J. (2016).
Deep Residual Learning for Image Recognition.
\emph{Proceedings of the IEEE Conference on Computer Vision and Pattern Recognition}, 770--778.
\href{https://doi.org/10.1109/CVPR.2016.90}{https://doi.org/10.1109/CVPR.2016.90}.

\bibitem[Vaswani et~al.(2017)]{Vaswani2017}
Vaswani, A., Shazeer, N., Parmar, N., Uszkoreit, J., Jones, L., Gomez, A. N., Kaiser, L., and Polosukhin, I. (2017).
Attention Is All You Need.
In \emph{Advances in Neural Information Processing Systems 30}, 5998--6008.
\href{https://arxiv.org/abs/1706.03762}{arXiv:1706.03762}.

\bibitem[Devlin et~al.(2019)]{Devlin2019}
Devlin, J., Chang, M.-W., Lee, K., and Toutanova, K. (2019).
BERT: Pre-training of Deep Bidirectional Transformers for Language Understanding.
In \emph{Proceedings of NAACL-HLT}, 4171--4186.
\href{https://doi.org/10.18653/v1/N19-1423}{https://doi.org/10.18653/v1/N19-1423}.

\bibitem[Gulshan et~al.(2016)]{Gulshan2016}
Gulshan, V., Peng, L., Coram, M., et~al. (2016).
Development and Validation of a Deep Learning Algorithm for Detection of Diabetic Retinopathy in Retinal Fundus Photographs.
\emph{JAMA}, 316(22), 2402--2410.
\href{https://doi.org/10.1001/jama.2016.17216}{https://doi.org/10.1001/jama.2016.17216}.

\bibitem[Esteva et~al.(2017)]{Esteva2017}
Esteva, A., Kuprel, B., Novoa, R. A., Ko, J., Swetter, S. M., Blau, H. M., and Thrun, S. (2017).
Dermatologist-level classification of skin cancer with deep neural networks.
\emph{Nature}, 542, 115--118.
\href{https://doi.org/10.1038/nature21056}{https://doi.org/10.1038/nature21056}.

\bibitem[Jumper et~al.(2021)]{Jumper2021}
Jumper, J., Evans, R., Pritzel, A., Green, T., Figurnov, M., Ronneberger, O., Tunyasuvunakool, K., Bates, R., Zidek, A., Potapenko, A., Bridgland, A., Meyer, C., Kohl, S. A. A., Ballard, A. J., Cowie, A., Romera-Paredes, B., Nikolov, S., Jain, R., Adler, J., Back, T., Petersen, S., Reiman, D., Clancy, E., Zielinski, M., Pacholska, M., Berghammer, T., Bodenstein, S., Silver, D., Vinyals, O., Senior, A. W., Kavukcuoglu, K., Kohli, P., and Hassabis, D. (2021).
Highly accurate protein structure prediction with AlphaFold.
\emph{Nature}, 596, 583--589.
\href{https://doi.org/10.1038/s41586-021-03819-2}{https://doi.org/10.1038/s41586-021-03819-2}.

\bibitem[Li et~al.(2024)]{Li2024NatureMedicine}
Li, J., Guan, Z., Wang, J., et~al. (2024).
Integrated image-based deep learning and language models for primary diabetes care.
\emph{Nature Medicine}, 30(10), 2886--2896.
\href{https://doi.org/10.1038/s41591-024-03139-8}{https://doi.org/10.1038/s41591-024-03139-8}.

\bibitem[Egghe(2006)]{Egghe2006}
Egghe, L. (2006).
Theory and practise of the g-index.
\emph{Scientometrics}, 69(1), 131--152.
\href{https://doi.org/10.1007/s11192-006-0144-7}{https://doi.org/10.1007/s11192-006-0144-7}.

\bibitem[Zhang(2009)]{Zhang2009}
Zhang, C.-T. (2009).
The e-index, complementing the h-index for excess citations.
\emph{PLOS ONE}, 4(5), e5429.
\href{https://doi.org/10.1371/journal.pone.0005429}{https://doi.org/10.1371/journal.pone.0005429}.

\bibitem[Hirsch(2019)]{Hirsch2019alpha}
Hirsch, J. E. (2019).
$h_{\alpha}$: An index to quantify an individual's scientific leadership.
\emph{Scientometrics}, 118(2), 673--686.
\href{https://doi.org/10.1007/s11192-018-2994-1}{https://doi.org/10.1007/s11192-018-2994-1}.




\end{thebibliography}
\end{document}